# Virtual-Thing: Thing Description based Virtualization

Hassib Belhaj Hassine
*Technical University of Munich*
hassib.belhaj@tum.de

Ege Korkan
*Technical University of Munich*
ege.korkan@tum.de

Sebastian Steinhorst
*Technical University of Munich*
sebastian.steinhorst@tum.de

*Abstract*— Integrating different Internet of Things devices from different manufacturers to create a mashup scenario can be a tedious and error prone task that involves studying non-standard datasheets. A Thing Description (TD) as defined by the World Wide Web Consortium (W3C) can make such a task less complicated by providing a standardized model for describing the metadata and the interface of a Web of Things (WoT) entity. However, a situation where a mashup developer has access to a Thing's TD before having access to the Thing itself may still arise. A way of simulating devices based only on their TDs is thus helpful during the development process of a mashup. In this work we present a method of creating a virtual Thing that simulates the behavior of a WoT-enabled entity based only on its Thing Description.

*Keywords— Web of Things, Thing Description, Virtualization, Virtual Thing, Internet of Things*

## I. INTRODUCTION

Internet connected devices and appliances are gaining widespread acceptance as part of the Internet of Things (IoT), yet most vendors rely on proprietary interfaces and non-standard software to manage them. This makes connecting multiple IoT devices in a mashup scenario very resource intensive and error prone. The Web of Things (WoT) is a solution to this problem. It consists of a set of approaches, software architectures and programming patterns intended to enable interoperability across IoT Platforms and application domains. As part of this effort, the World Wide Web Consortium (W3C) has recently launched the WoT Working Group to work on creating a set of standards for the Web of Things. One such standard, that is currently close to being finalized, is the WoT Thing Description standard.

A Thing Description (TD) describes the metadata and interfaces of a Thing, where a Thing is an abstraction of a physical or virtual entity that provides interactions to and participates in the Web of Things [1]. Every WoT entity can use its TD as a standard way of presenting itself and telling other entities how to communicate with it and how to correctly invoke its functionality. This makes creating a mashup scenario where multiple devices from multiple vendors are connected to each other very easy, removing the need for the developer to go through a multitude of differently formatted data sheets to understand how to communicate with each device specifically. The resulting complexity reduction can result in significant time savings and reduce the programming error rates significantly.

While integrating multiple WoT devices is relatively easy based on their TDs, programmers might still find themselves in a situation where they have to start working on the software to control or access a certain device based only on its TD, without getting full access to the device itself during the development phase. This makes testing very difficult. Having the ability to simulate a Thing based only on its TD is thus very helpful. In Section II, we present the essential building blocks of the Web of Things, including Thing Descriptions and the *node-wot* reference implementation. In Section III, we then discuss an approach to create a virtual instance of a Thing, based on its TD. The virtual Thing has exactly the same interface as the simulated Thing, and can be used for testing and integration purposes in case access to the original instance is not available.

The limitations of our approach and potential future work are discussed in Section IV and Section V presents a conclusion.

## II. THE WEB OF THINGS BUILDING BLOCKS

### A. WoT Architecture: Thing Descriptions

The Web of Things started as an academic initiative with the main goals of enabling interoperability between different IoT devices and platforms as well as improving their usability. In 2016, the W3C started the WoT working group to define a set of standard mechanisms to describe IoT interfaces and allow IoT devices to easily communicate with each other independently from their underlying implementations [2]. The first set of WoT building blocks is now being standardized and includes the WoT Thing Description, and WoT Architecture. Of these, the WoT Thing Description is the primary and most important building block, describing the public interface of a Thing. [2]

Thing Descriptions use a predefined vocabulary to describe the set of possible interactions and functionalities provided by an IoT device and be able to access them. This makes it possible to integrate multiple devices without needing to know their underlying implementation, allowing diverse applications to be interoperable. As an example, a smart air conditioner can provide a TD that describes all the interactions it supports, making it possible for a smart thermostat to understand how to communicate with it and control it just by knowing its TD, even if they are from different manufacturers. If the thermostat also provides a TD, they can then both be accessed and controlled by an authorized smart controller that might also control a multitude of other Things in a smart home. This can significantly reduce the integration efforts and bug frequency when creating systems that are composed of multiple IoT devices. Fig. 1 shows an example of how this can work.

The data contained in a TD includes the name of the Thing, its unique identification, its security requirements, an optional human readable description, and all the possible interactions it supports [1]. These are divided in three types: Properties, Actions and Events:

*a) Properties* are values that expose the internal state of the Thing. They can be retrieved and optionally modified. They may also be observable.

*b) Actions* on the other hand represent the functions of the Thing. They may change its internal state or the values of its properties.

*c) Events* represent state transitions, and are asynchronously pushed by the Thing to subscribed observers.

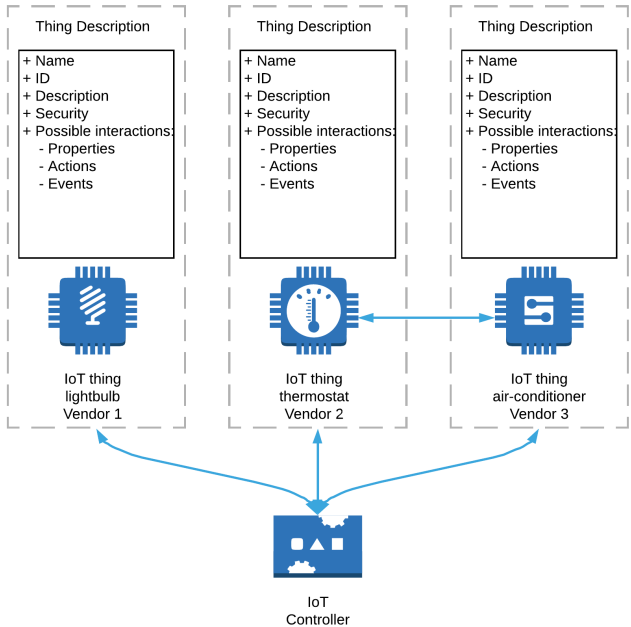

Fig. 1. An example of how multiple Things from different vendors can interact with each other, or be controlled from an IoT controller, based only on the information provided in their Thing Descriptions

List. 1 shows an example of how a TD for coffee machine might look like. In addition to the machine's name, ID and description, the TD presents the property **state**, which can have a value of either **ready**, **brewing** or **error** and can be accessed by adding the URL given in its forms field to the base URL. It also presents the action "brew", which receives a string as input and the event error which can be generated by the coffee machine and send a string to all observers.

TDs can also be made machine readable, using the JSON-LD format to extended them with semantic annotations and contextual models [5].

### B. Example Implementation: node-wot

One way of creating WoT compatible Things and of parsing TDs is to use a WoT framework such as *node-wot*.

*Node-wot* is a reference implementation of multiple proposed WoT standards. It is a part of the Eclipse Thingweb project [6] under the direction of the Eclipse Foundation and provides a WoT Thing Description parser and serializer, as well as several protocol implementations based on the WoT Binding Templates [3] and a programming runtime based on the WoT Scripting API [4].

*Node-wot* provides a server/client (servient) architecture that makes it possible to create a Thing from scratch with very little programming work, or to consume a TD and easily communicate with its corresponding Thing. It can be as a standalone application or as a Node.js module which is the approach that we use in this work. We import the node-wot module and extend it to create the Virtual Thing generator discussed in Section III. Node-wot is written in Typescript and needs to be compiled to JavaScript before being run by Node.js making Typescript the ideal language for extending it. Things created with node-wot automatically have a TD generated for them and can use any of the protocol bindings provided by the node-wot team, including for HTTP, HTTPS, MQTT and COAP. It also handles the underlying communication overhead when communicating with other WoT Things, providing an easy to use WoT Scripting API compatible programming interface.

---
[1] https://www.npmjs.com/package/virtual-thing

```
01  {
02      "@context": "https://www.w3.org/2019/wot/td/v1",
03      "id": "urn:dev:org:esitum-CoffeeMachine-001",
04      "title": "Coffee-Machine",
05      "description": "A WoT enabled coffee machine",
06      "security": ["no"]
07      "securityDefinitions":{"no":{"scheme":"nosec"}},
08      "base": "http://10.0.0.1/coffee-machine",
09      "properties": {
10          "state": {
11              "type": "string",
12              "enum": ["Ready", "Brewing", "Error"],
13              "forms": [{"href": "/properties/state"}]
14          }
15      },
16      "actions": {
17          "brew": {
18              "input": {
19                  "type": "string",
20                  "enum": ["espresso", "cappuccino"]
21              },
22              "forms": [{"href": "/actions/brew"}]
23          }
24      },
25      "events": {
26          "error": {
27              "data": {"type": "string"},
28              "forms": [{"href": "/events/error"}]
29          }
30      }
31  }
```

List. 1. An example Thing Description of WoT enabled coffee machine

Due to the multiple WoT standards being still in draft mode and undergoing rapid changes, we limit our protocol support to the HTTP bindings in this work. It is worth noting that any other protocol supported by *node-wot* can easily be used instead.

### III. THING VIRTUALIZATION

#### A. Virtual-Thing: Concept

Often, a developer or a system integrator might find himself in a situation where the interface for a Thing has been defined well before the hardware is available. This means that a TD exists, and work on integrating or extending the Thing can begin. However, testing in this situation is very difficult, and having a Virtual Thing with the same interface to act as a dummy can be very helpful. This is for example the case if one purchases a WoT enabled Thing online. Its TD can be available in an online repository, allowing work to start on integrating it with other Things as part of a big system. In such a situation, it might still be necessary to wait for the physical device to be shipped before starting to test the code. To help solve this problem, we propose virtual-thing [1], a method and a program for virtualizing a Thing based only on its TD.

A WoT Thing provides three distinct types of interactions: properties, actions and events. To simulate an interaction, we use its description from the TD as a JSON Schema and generate random data that conforms to it. The data itself might be random, but its format is the same as described in the TD. This means that generated responses will always have the correct type, and that a generated object will always have all the properties described in the TD, even if their value is random. Other parts of the schema such as *enum* are also respected. As an example, if a property of type string has an *enum* field attached to it, virtual-thing will always generate a string that is one of the enumerated values. In the case of the property "state" seen in Listing 1, the response will randomly switch between "Ready", "Brewing" and "Error".

Properties, as described in a TD, do not separate the information about their data format from the rest of their meta-data. This makes simulating them slightly more complicated than for actions or events, as the Virtual Thing has to parse the TD and filter the JSON Schema-relevant information first. Table 1 shows the relevant properties for each response type.

Actions on the other hand have an input and an output property which already contain a schema describing the data they receive or return. Virtual-thing uses this property to validate the received action inputs and to return random data that still conforms to the JSON Schema described in the output property.

Events have a property called data that is also a JSON Schema representing the data that they emit. This is used by virtual-thing to generate corresponding data. The program can be configured to either not generate any event, automatically generate events with a random interval that is automatically selected in the range of 5 to 60 seconds, or to generate events after a user configured set interval.

The resulting architecture is shown in Figure 2.

TABLE I. JSON SCHEMA RELEVANT PROPERTIES BY TYPE

| Property type | Relevant keys |
|---|---|
| -- All types -- | `type, enum, const, oneOf` |
| integer / number | `minimum, maximum` |
| array | `items, minItems, maxItems` |
| object | `properties, required` |

*B. Virtual-Thing: Implementation*

Virtual-thing is a Node.js program that uses *node-wot* as library and extends it. It is written in typescript and uses the fundamental module structure of Node.js. It can thus either be used as a standalone application or be imported as library. It works by getting a Thing Description as an input parameter, and may also get an existing *node-wot* servient as input. If no servient is provided, one is automatically started. The Virtual Thing then attaches itself to the servient and presents itself by generating a TD that is an exact copy of the input TD, with the exception of the forms URLs being changed to reflect the fact that the Virtual Thing is running on a different address.

The IP address used to run the virtual Thing can either be set by the Servient in case virtual-thing is used as a library, or be fixed in a configuration file in case virtual-thing is used as a standalone application. The HTTP port can also be configured in the same way.

IV. LIMITATIONS AND FUTURE WORK

The virtual-thing software module has some known limitations, most of them due to limitation in the current implementation of the underlying *node-wot* framework. One such limitation is that the form URLs cannot be arbitrary and have to follow the convention set by *node-wot*. Properties are always accessible under:

`<IP>/<thing-name>/properties/<property-name>`

The URL of actions and events also follow a similar convention. This limitation does however not affect clients programmed using the scripting API, as they do not directly access the URLs.

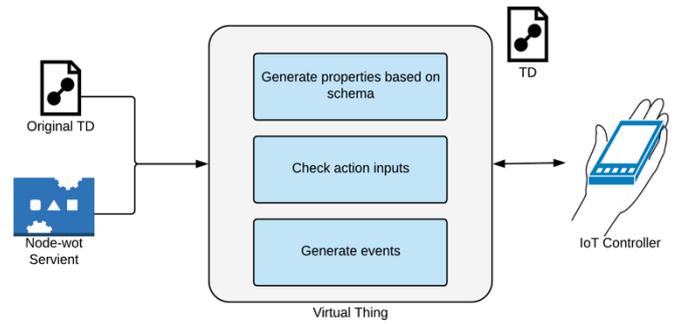

Fig. 2. To start a Virtual Thing and attach it to an existing node-wot servient, only a TD is needed. The Virtual Thing automatically generates data for the interactions and creates a new TD that is identical to the original one, except for changed URLs to reflect that is running locally.

Other limitations include:

- It is not possible to correctly reject property writes with an error message
- It is not possible to reject an event subscription with an error if it has the wrong format
- Event subscription / cancelation data is not passed on
- No way to subscribe to properties
- Node-wot only supports JSON as communication payload format

Solving these problems by changing *node-wot*, as well as improving support for protocols other than HTTP in the virtual-thing module are important subjects for future work.

V. CONCLUSION

In this paper we introduced virtual-thing, a method of simulating a WoT Thing based only on its Thing Description using a *node-wot* based program. We discussed how the functionalities provided by virtual-thing facilitate the development and integration work in situations where a developer might have access to a TD before getting access to the Thing itself, making it possible to test code with a virtual entity that has the same interface as the object being tested. Finally, we described some of the limitations of the current implementation.

REFERENCES

[1] S. Kaebisch and T. Kamiya, Web of Things (WoT) Thing Description - W3C Working Draft 21 October 2018. https://www.w3.org/TR/2018/WD-wot-thing-description-20181021/

[2] M. Kovatsch, K. Kajimoto, R. Matsukura, M. Lagally and T. Kawaguchi, Web of Things (WoT) Architecture - W3C Editor's Draft 07 February 2019. https://www.w3.org/TR/2017/WD-wot-architecture-20170914/

[3] M. Koster, Web of Things (WoT) Protocol Binding Templates - W3C Working Group Note 5 April 2018 https://www.w3.org/TR/2018/NOTE-wot-binding-templates-20180405/

[4] Z. Kis, K. Nimuraand and D. Peintner, Web of Things (WoT) Scripting API - W3C Working Draft 29 November 2018 https://www.w3.org/TR/2018/WD-wot-scripting-api-20181129/

[5] S. Kaebisch and D. Anicic., 2016. Thing description as enabler of semantic interoperability on the Web of Things. In Proc. IoT Semantic Interoperability Workshop (pp. 1-3).

[6] M. Kovatsch, D. Peintner, C. Glomb et al., Eclipse Thingweb https://projects.eclipse.org/projects/iot.thingweb